\documentstyle[12pt,epsf,twoside]{article}

\advance\textheight by \topskip
\newsavebox{\Forh}
\newsavebox{\Abox}

  \newlength{\absize}
  \renewcommand{\baselinestretch}{2.0}
  \renewcommand{\arraystretch}{2.0}
  
\begin{document}
  \thispagestyle{empty}
  \pagestyle{empty}
  \renewcommand{\thefootnote}{\fnsymbol{footnote}}
  \newcommand{\starttext}{\newpage\normalsize
    \pagestyle{plain}
    \setlength{\baselineskip}{4ex}\par
    }
\newcommand{\preprint}[1]{
  \begin{flushright}
    \setlength{\baselineskip}{3ex} #1
  \end{flushright}}
\renewcommand{\title}[1]{
  \begin{center}
    \LARGE #1
  \end{center}\par}
\renewcommand{\author}[1]{
  \vspace{2ex}
  {\Large
   \begin{center}
     \setlength{\baselineskip}{3ex} #1 \par
   \end{center}}}
\renewcommand{\thanks}[1]{\footnote{#1}}
\renewcommand{\abstract}[1]{
  \vspace{2ex}
  \normalsize
  \begin{center}
    \centerline{\bf Abstract}\par
    \vspace{2ex}
    \parbox{\absize}{#1\setlength{\baselineskip}{2.5ex}\par}
  \end{center}}

\setlength{\parindent}{3em}
\setlength{\footnotesep}{.6\baselineskip}
\newcommand{\myfoot}[1]{
  \footnote{\setlength{\baselineskip}{.75\baselineskip}#1}}
\renewcommand{\thepage}{\arabic{page}}
\setcounter{bottomnumber}{2}
\setcounter{topnumber}{3}
\setcounter{totalnumber}{4}
\newcommand{\figsize}{}
\renewcommand{\bottomfraction}{1}
\renewcommand{\topfraction}{1}
\renewcommand{\textfraction}{0}

 \newlength{\rtcol}
 \setlength{\rtcol}{\tabcolsep}

\preprint{University of Bergen, Department of Physics \\
Scientific/Technical Report No.\ 1993-02 \\ ISSN~0803-2696}

\newlength{\mm}
\newlength{\mmh}
\newlength{\ltT}
\newlength{\gmljot}
\setlength{\gmljot}{\jot}
\newcommand{\ko}{\dagger}
\newcommand{\st}[1]{{\scriptstyle{#1}}}
\newcommand{\what}[3]{\settowidth{\ltT}{\Dis{#3}}
              \makebox[\ltT]{$\rule{#2\mmh}{0mm}\widehat{\makebox[#1\mm]
              {\Dis{#3\rule{#2\mm}{0mm}}}}$}}
\newcommand{\wtilde}[3]{\settowidth{\ltT}{\Dis{#3}}
            \makebox[\ltT]{$\rule{#2\mmh}{0mm}\widetilde{\makebox[#1\mm]
              {\Dis{#3\rule{#2\mm}{0mm}}}}$}}
\newcommand{\Ab}{A_{{b}}}
\newcommand{\At}{A_{{t}}}
\newcommand{\Ad}{A_{{d}}}
\newcommand{\Au}{A_{{u}}}
\newcommand{\Ae}{A_{{e}}}
\newcommand{\as}{\alpha_{s}}
\newcommand{\al}{\alpha}
\newcommand{\alep}{\tilde{{l}}}
\newcommand{\aeep}{\tilde{{e}}}
\newcommand{\alp}{\dot{\alpha}}
\newcommand{\be}{\beta}
\newcommand{\Biiggls}[2]{\hspace*{-1mm}
   \settowidth{\unitlength}{{$\displaystyle
                               \left#2\rule{0mm}{#1}\right.$}}
         \addtolength{\unitlength}{0.84823mm}
         \makebox[0mm]{ \rule{0mm}{#1}
         \begin{picture}(0,0)
                      \put(0,0){$\left#2\rule{0mm}{#1}\right.$}
                      \end{picture}}\hspace*{\unitlength}}
\newcommand{\Biiggrso}[3]{\hspace*{-2.5mm}
    \settowidth{\unitlength}{{$\displaystyle
                    \left.\rule{0mm}{#1}\right#2$}}
                         \addtolength{\unitlength}{1.5mm}
                         \makebox[0mm]{ \rule[- #1]{0mm}{#1}
                         \begin{picture}(0,0)
                          \put(0,0){$\left.\rule{0mm}{#1}\right#2^{#3}$}
                         \end{picture}}\hspace*{\unitlength}}
\newcommand{\dC}{\delta_{{{C}}}}
\newcommand{\Del}{\Delta}
\newcommand{\Dis}[1]{$\displaystyle #1$}
\newcommand{\Disp}[1]{{\displaystyle #1}}
\newcommand{\eps}{\epsilon}
\newcommand{\equ}{e_{{u}}}
\newcommand{\eq}{e_{q}}
\newcommand{\Ga}{\Gamma}
\newcommand{\GeV}{\mbox{{\rm GeV}}}
\newcommand{\ga}{\gamma}
\newcommand{\Hns}{{H}^{0}}
\newcommand{\hnl}{{h}^{0}}
\newcommand{\hA}{{A}}
\newcommand{\Hl}{{H}}
\newcommand{\hl}{h_{{l}}}
\newcommand{\huq}{h_{{u}}}
\newcommand{\hdq}{h_{{d}}}
\newcommand{\hc}{\mbox{{\rm h.c.}}}
\newcommand{\jots}[1]{\setlength{\jot}{#1 mm}}
\newcommand{\jotr}{\setlength{\jot}{\gmljot}}
\newcommand{\la}{\lambda}
\newcommand{\La}{\Lambda}
\newcommand{\Mb}{m_{{b}}}
\newcommand{\Mt}{m_{{t}}}
\newcommand{\Md}{m_{{d}}}
\newcommand{\Mu}{m_{{u}}}
\newcommand{\Msq}{m_{\tilde{q}}}
\newcommand{\Msba}{m_{\tilde{{b}}1}}
\newcommand{\Msbb}{m_{\tilde{{b}}2}}
\newcommand{\Msta}{m_{\tilde{{t}}1}}
\newcommand{\Mstb}{m_{\tilde{{t}}2}}
\newcommand{\MsT}{\wtilde{3}{0.2}{m}_{T}}
\newcommand{\MsB}{\wtilde{3}{0.2}{m}_{B}}
\newcommand{\MsQT}{\wtilde{3}{0.8}{M}_{\!T}}
\newcommand{\Msu}{\wtilde{3}{0.2}{m}_{U}}
\newcommand{\Msd}{\wtilde{3}{0.2}{m}_{\!D}}
\newcommand{\MsQU}{\wtilde{3}{0.8}{M}_{\hspace*{-1mm}U}}
\newcommand{\MsQ}{\wtilde{3}{0.8}{M}_{\!Q}}
\newcommand{\mA}{m_{{A}}}
\newcommand{\MC}{m_{{\chi_{1}}}}
\newcommand{\mC}{m_{{\chi_{2}}}}
\newcommand{\mh}{m_{{h}^{0}}}
\newcommand{\mW}{m_{{W}}}
\newcommand{\mZ}{m_{{Z}}}
\newcommand{\me}{m_{{e}}}
\newcommand{\MMse}{\wtilde{3}{0.8}{M}_{\hspace*{-1.2mm}E}}
\newcommand{\MHe}{M_{\!H_{1}}}
\newcommand{\MHt}{M_{\!H_{2}}}
\newcommand{\Msle}{\wtilde{3}{0.2}{m}_{\!E}}
\newcommand{\MNS}{m_{\Disp{\tilde{{\scriptstyle \La}}}}}
\newcommand{\mlN}{m_{\Disp{\tilde{{\scriptstyle \la}}}}}
\newcommand{\ned}[1]{\makebox[0mm]{$\rule{0mm}{#1\mm}$}}
\newcommand{\och}{\overline{\chi}}
\newcommand{\ola}{\overline{\lambda}}
\newcommand{\oLa}{\overline{\Lambda}}
\newcommand{\othth}{\overline{\theta}\,\overline{\theta}}
\newcommand{\oxi}{\overline{\xi}}
\newcommand{\pad}{\partial}
\newcommand{\pC}{\phi_{{{C}}}}
\newcommand{\sba}{\tilde{{b}}_{1}}
\newcommand{\sbb}{\tilde{{b}}_{2}}
\newcommand{\squ}{\tilde{q}}
\newcommand{\suq}{\tilde{{u}}}
\newcommand{\scq}{\tilde{{c}}}
\newcommand{\sdq}{\tilde{{d}}}
\newcommand{\ssq}{\tilde{{s}}}
\newcommand{\stq}{\tilde{{t}}}
\newcommand{\sta}{\tilde{{t}}_{1}}
\newcommand{\stb}{\tilde{{t}}_{2}}
\newcommand{\si}{\sigma}
\newcommand{\ssst}[1]{{\scriptscriptstyle{#1}}}
\newcommand{\tauLq}{\tau_{\raisebox{-0.3mm}{$\st{\hspace*{-0.2mm}
               \squ^{\hspace*{-0.2mm}\raisebox{-0.5mm}{$\ssst{L}$}}}$}}}
\newcommand{\tauLl}{\tau_{\raisebox{-1mm}{$\st{\hspace*{-0.2mm}
               \alep^{\hspace*{-0.1mm}\raisebox{0.3mm}{$\ssst{L}$}}}$}}}
\newcommand{\tC}{\theta_{{{C}}}}
\newcommand{\Tsp}{\mbox{\scriptsize T}}
\newcommand{\thth}{\theta\theta}
\newcommand{\TR}{\mbox{Tr}}
\newcommand{\tb}{\theta_{\tilde{{b}}}}
\newcommand{\tht}{\theta_{\tilde{{t}}}}
\newcommand{\tW}{\theta_{\mbox{{\tiny W}}}}
\newcommand{\Wpm}{{W}^{\pm}}

\vfill

\title{Effects of scalar mixing in ${g}{g} \rightarrow \mbox{Higgs}
       \rightarrow \ga\ga$}

\vfill

\author{Bjarte Kileng \\ \hfil \\
        Department of Physics\thanks{electronic mail address:
                {\tt kileng@vsfys1.fi.uib.no}}\\
        University of Bergen \\ All\'egt.~55, N-5007 Bergen, Norway }
\date{}

\vfill \abstract{ We discuss the effects of mixing of scalars belonging
to left-- and right--chiral MSSM super--multiplets on the process
${g}{g}\rightarrow \hnl \rightarrow \ga\ga$,
which is a promising channel for the discovery of the lightest CP--even
MSSM Higgs boson at the LHC and the SSC. The effects of the left--right
scalar mixing are small in most of parameter space. However, there
exist regions where the mixing is important. Due to mixing, there can
be large enhancements of the cross section in a region of parameter
space where the no--mixing approximation yields a value that is too low
for experimental discovery of the $\hnl$ at the LHC and the SSC. }

\vfill

 \starttext

\section{Introduction}

The Standard Model can only be an effective low--energy theory which is
replaced at some scale by a more fundamental theory~\cite{Hab}. One of
the most studied and promising alternative theories is the Minimal
Supersymmetric Standard Model (MSSM)~\cite{Hab,MuKi,BaWe,Nill,Barb}.
The MSSM introduces many fields and parameters which are not present in
the standard model.

\medskip

Supersymmetric theories with supersymmetry effectively broken in the
vicinity of the electroweak scale can naturally accomodate elementary
Higgs bosons without introducing quadratic divergences. The bosonic
Higgs sector of the MSSM comprises the CP--even fields $\hnl$, $\Hns$,
the CP--odd field $\hA$ and the charged field $\Hl$. At tree--level, in
the MSSM, the lightest CP--even Higgs boson ($\hnl$) will be lighter
than the ${Z}^{0}$. But the masses of the MSSM Higgs bosons are subject
to large radiative corrections associated with the top quark and its
SU(2) and supersymmetric partners~\cite{OYY,ERZF,HEHRH}. As a result, a
significant region of parameter space for the MSSM Higgses can be
beyond the reach of LEP 200~\cite{ZKFZ}. The discovery of MSSM Higgs
bosons will then have to wait for the LHC and/or the SSC. Even if a
neutral Higgs boson is found at LEP, the LHC and/or the SSC will be
needed in order to determine whether it is a MSSM Higgs, and to
investigate its properties.

\medskip

At the LHC and the SSC the dominant production mechanism for the
neutral Higgs bosons will be gluon fusion~\cite{ZKFZ},
${g}{g}\rightarrow \hnl,\Hns,\hA$. For the lightest CP--even Higgs
boson~($\hnl$), a promising decay channel will be the two--photon
decay, $\hnl\rightarrow \ga\ga$. The signal is clean, and the decay
rates are high enough for detection in large regions of parameter
space. Both the gluon fusion and the two--photon decay are sensitive to
the superparticles (coloured for gluon fusion and electrically charged
for two--photon decay) which get
masses\myfoot{ The squarks, sleptons, charginos and scalar Higgses get
contributions to their masses from both the Higgs mechanism and the
soft breaking terms of eq.~(\ref{JKhy}) below.}
from the Higgs mechanism.

\medskip

In the MSSM each quark and lepton has two superpartners. They
correspond to the left-- and right--chiral states of the quarks and
leptons. We will refer to these superpartners as left-- and right
scalars. The physical mass eigenstates are in general a superposition
of the left-- and right--scalar states. In this paper we will calculate
the effect of this mixing on the process \begin{equation}
{g}{g}\longrightarrow \hnl \longrightarrow \ga\ga \hspace*{1mm}.
\label{uiJKty} \end{equation} We shall describe this as a succession of
two independent processes, ${g}{g}\rightarrow \hnl$ followed by $\hnl
\rightarrow \ga\ga$. Apart from phase space factors, the cross section
for the former process is given by the decay rate for the inverse
process, $\Ga^{{g}{g}}=\Ga(\hnl \rightarrow {g}{g})$, while the latter
is given by $\Ga^{\ga\ga}=\Ga(\hnl \rightarrow \ga\ga)$. Both of these
decay rates are affected by the mixing of the left-- and right--scalar
states. In order to study the effect of the left--right scalar mixing
on the process~(\ref{uiJKty}), we consider the quantity
\begin{equation} F =
\frac{\Ga^{{g}{g}}_{ \mbox{{ \footnotesize mixing}}}}{\Ga^{{g}{g}}_{
                                     \mbox{{ \footnotesize no mixing}}}}
\times
\frac{\Ga^{\ga\ga}_{\mbox{{ \footnotesize mixing}}}}{\Ga^{\ga\ga}_{
                                     \mbox{{ \footnotesize no mixing}}}}
        \hspace*{1mm}.
        \label{OiuY} \end{equation}

\medskip

The gluon fusion (or two--gluon decay which we consider in the
calculations) and two--photon decay of the neutral MSSM Higgs bosons
are due to loop diagrams. The one--loop decay of the CP--even Higgs
bosons into two photons is due to quarks, squarks, leptons, sleptons,
$\Wpm$, charginos and charged Higgses. The one--loop decay of the
CP--even Higgs bosons into two gluons is due to quarks and squarks. The
dominant contribution in the quark sector (including the lepton sector
in the case of two--photon decay) comes from the top and bottom quarks.
For some regions in parameter space, the one--loop contribution from
the bottom quark may be comparable, or even larger, than that due to
the top quark.

\medskip

This paper is organized as follows: In sect.~\ref{inms} we discuss
general properties of the MSSM. The parameters and fields of the MSSM
are introduced. In sect.~\ref{partcont} we study the particle content
relevant for  gluon fusion and two--photon decay of the lightest
CP--even Higgs boson $\hnl$. In sect.~\ref{dewid} these results enable
us to obtain the two--photon and two--gluon decay widths of the
lightest CP--even Higgs boson $\hnl$, with and without left--right
scalar mixing. In sect.~\ref{numd} we discuss numerical results showing
effects of the left--right scalar mixing on the process
${g}{g}\rightarrow \hnl \rightarrow \ga\ga$. Sect.~\ref{conc} contains
our conclusions.

\section{The MSSM Lagrangian density}

\label{inms}

We shall in this section give a brief discussion of the various parts
of the MSSM Lagrangian density. The different fields and some other
notations are given in table~\ref{kjtyFG}.

\medskip

The Lagrangian density for the MSSM is given by
\begin{equation} {\cal L} = {\cal L}_{\mbox{{\scriptsize SU(3)}}}
 + {\cal L}_{\mbox{{\scriptsize SU(2)$\times$U(1)}}}
 + {\cal L}_{\mbox{{\scriptsize Soft}}} \hspace*{1mm}. \end{equation}
We will ignore inter--generational mixing. The SU(2)$\times$U(1) part
(the hat refers to superfields) is given by~\cite{BaWe}
\jots0
\begin{eqnarray} {\cal L}_{\mbox{{\scriptsize SU(2)}}\times
    \mbox{{\scriptsize U(1)}}} & = & \Biggl\{ \biggl[
    \frac{1}{8g^{2}}\TR\left[ W^{\al}W_{\al} \right]
    + \frac{1}{4}w^{\al}w_{\al}
    - \mu \what{3}{0.7}{H}^{\Tsp}_{1} \eps \what{3}{0.7}{H}_{2}
    \label{hhcoup} \\*
& & \hspace*{15mm} \mbox{}
    + \hl\what{3}{0.7}{L}^{\Tsp}\eps \what{3}{0.7}{H}_{1}
      \what{3}{0.7}{E}^{R}
    + \hdq\what{3}{0.7}{Q}^{\Tsp}\eps\what{3}{0.7}{H}_{1}
      \what{3}{0.7}{D}^{R}
    + \huq\what{3}{0.7}{Q}^{\Tsp}\eps\what{3}{0.7}{H}_{2}
      \what{3}{0.2}{U}^{R}
 \biggr]_{\thth}
 + \hc  \Biggr\} \nonumber \\[5mm]
& & + \Biggl[ \what{3}{0.7}{L}^{\ko}e^{\left( 2gV - g'v \right)}
      \what{3}{0.7}{L}
    + \what{3}{0.7}{E}^{R\ko}e^{2g'v}\what{3}{0.7}{E}^{R}
    + \what{3}{0.7}{Q}^{\ko}e^{(2gV+g'v/3)}\what{3}{0.7}{Q}
    + \what{3}{0.2}{U}^{R\ko}e^{-4g'v/3}\what{3}{0.2}{U}^{R}
      \nonumber \\*
& & \hspace*{15mm} \mbox{}
    + \what{3}{0.7}{D}^{R\ko}e^{2 g'v/3}\what{3}{0.7}{D}^{R}
    + \what{3}{0.7}{H}^{\ko}_{1}e^{ (2gV-g'v) }\what{3}{0.7}{H}_{1}
    + \what{3}{0.7}{H}^{\ko}_{2}e^{ (2gV+g'v) }\what{3}{0.7}{H}_{2}
      \Biggr]_{\thth\othth} \nonumber \hspace*{1mm},
\end{eqnarray}
and the SU(3) part is given by
\begin{equation} {\cal L}_{\mbox{{\scriptsize SU(3)}}} =
   \frac{1}{8g_{S}^{2}}\TR\left[ W_{S}^{\al}W_{S\al} \right]_{\thth}
 + \left[ \what{3}{0.7}{Q}^{L\ko}_{\mbox{\scriptsize SU(3)}}e^{2g_{S}
   V_{S}}\what{3}{0.7}{Q}^{L}_{\mbox{\scriptsize SU(3)}}
  + \what{3}{0.7}{Q}^{R}_{\mbox{\scriptsize SU(3)}}e^{-2g_{S}V_{S}}
    \what{3}{0.7}{Q}^{R\ko}_{\mbox{\scriptsize SU(3)}}
    \right]_{\thth\othth} \hspace*{1mm}. \end{equation}
The gauge--invariant (soft) supersymmetry breaking terms are, in terms
of component fields~\cite{GiGr}
\begin{eqnarray} {\cal L}_{\mbox{{\scriptsize Soft}}} & = & \Biggl\{
      \be^{Hh} H_{1}^{\Tsp} \eps H_{2}
    + \frac{g\me\Ae}{\sqrt{2}\;\mW\cos\be}E^{\Tsp}\eps H_{1}\aeep^{R}
      \label{JKhy} \\*
& & \hspace*{15mm} \mbox{} + \frac{g\Md\Ad}{\sqrt{2}\;\mW\cos\be}
    Q^{\Tsp}\eps H_{1}\sdq^{R}
    - \frac{g\Mu\Au}{\sqrt{2}\;\mW\cos\be}Q^{\Tsp}\eps H_{2}\suq^{R}
    + \hc \Biggr\} \nonumber \\[7mm]
& & -  \MMse^{2}L^{\ko}L - \Msle^{2}\aeep^{R\ko}\aeep^{R}
    - \MsQU^{2}Q^{\ko}Q - \Msu^{2}\suq^{R\ko}\suq^{R}
    - \Msd^{2}\sdq^{R\ko}\sdq^{R} \nonumber \\[3mm]
& & - \MHe^{2} H_{1}^{\ko}H_{1} - \MHt^{2} H_{2}^{\ko}H_{2}
    + \frac{\mlN}{2}\left\{\la\la +\ola\;\ola\right\}
    + \frac{\MNS}{2} \sum_{I=1}^{3} \left\{\La_{I}\La_{I}
    + \oLa_{I}\oLa_{I}\right\} \hspace*{1mm}. \nonumber
\end{eqnarray}
\jotr
The auxiliary fields can be removed by means of the Euler--Lagrange
equations. With the vector superfields expressed in the Wess--Zumino
gauge (see~\cite{MuKi} or~\cite{BaWe}), the photon, the ${Z}^{0}$ and
the $\Wpm$ can be introduced as linear combinations of the SU(2) and
U(1) gauge fields. In order to determine the couplings of the MSSM, the
scalar Higgs sector must be expressed in terms of the vacuum
expectation values, the physical Higgs bosons and the Goldstone fields.
Furthermore, all mass matrices must be diagonalized.

\section{Particle content}

\label{partcont}

In the MSSM, to lowest order, the process ${g}{g} \rightarrow \hnl
\rightarrow \ga\ga$ gets contributions from diagrams containing fields
that are not present in the standard model, i.e. fields corresponding
to the super--particles and the charged Higgs boson.

\medskip

The CP--even Higgs sector is studied for instance in ref.~\cite{ERZ}
where the one--loop radiative corrections to the masses of the CP--even
Higgs bosons are calculated using the effective potential approach.
These corrections are included in our calculations of the two--photon
and two--gluon decay of the lightest CP--even MSSM Higgs boson. The
mixing angle $\al$ of the CP--even Higgs sector is then calculated
using the corrected mass matrix.

\medskip

The charged Higgs sector is studied for instance in ref.~\cite{BEZ}
where the one--loop radiative corrections to the mass of the charged
Higgs boson are calculated using the effective potential approach.
These corrections can be large in some regions of parameter space. We
therefore include them in our calculations.

\subsection{The slepton and squark sector}

 From the Lagrangian density of the MSSM we find the mass terms for the
squarks and sleptons. Neglecting inter--generational mixing, we get for
an up--type squark
\renewcommand{\arraystretch}{1}
\begin{equation} - \left( \begin{array} {c}
                 \suq^{L} \\
                 \suq^{R\ko}
           \end{array} \right)^{\ko}
        {\cal M}_{\suq} \left( \begin{array} {c}
                                         \suq^{L} \\
                                         \suq^{R\ko}
                    \end{array}    \right) \hspace*{1mm}. \end{equation}
At tree level, the mass matrix is given by
\renewcommand{\arraystretch}{0.8}
\begin{equation} {\cal M}_{\suq} = \left( \begin{array} {cc}
       \begin{array}{ll} - \frac{1}{6}\mZ^{2}\cos(2\be)
             + \frac{2}{3}\mW^{2}\cos(2\be) \\
                          \hspace*{4mm} +\Mu^{2} + \MsQU^{2} \end{array}
                              & \Mu\left\{ \Au + \mu\cot\be \right\} \\
        \Mu\left\{ \Au + \mu\cot\be \right\}
            & \ned{10} \begin{array}{ll} \frac{2}{3}\mZ^{2}\cos(2\be)
                  - \frac{2}{3}\mW^{2}\cos(2\be) \\
                          \hspace*{4mm} + \Mu^{2} + \Msu^{2} \end{array}
\end{array} \right) \hspace*{1mm}, \label{uptypq} \end{equation}
where $\MsQU$ and $\Msu$ are explicit mass terms of
${\cal L}_{\mbox{{\scriptsize Soft}}}$, and the off--diagonal terms
arise from the Higgs--Higgs coupling of eq.~(\ref{hhcoup}) and the
Yukawa couplings of ${\cal L}_{\mbox{{\scriptsize Soft}}}$. The
no--mixing case corresponds to
\begin{equation} \Au=\mu=0 \hspace*{1mm}. \label{iodhjGH} \end{equation}
Similarly, for a down--type squark we find
\begin{equation} {\cal M}_{\tilde{d}} = \left( \begin{array} {cc}
        \begin{array}{ll} - \frac{1}{6}\mZ^{2}\cos(2\be)
     - \frac{1}{3}\mW^{2}\cos(2\be) \\ \hspace*{4mm} + \Md^{2}
     + \MsQU^{2} \end{array}
                 & \Md\left\{ \Ad + \mu\tan\be \right\} \\
        \Md\left\{ \Ad + \mu\tan\be \right\}
          & \ned{10} \begin{array}{ll} - \frac{1}{3}\mZ^{2}\cos(2\be)
            + \frac{1}{3}\mW^{2}\cos(2\be) \\
                         \hspace*{4mm} + \Md^{2} + \Msd^{2} \end{array}
\end{array} \right) \hspace*{1mm}. \label{downtypq} \end{equation}
For a  charged slepton, the mass matrix is given by
\renewcommand{\arraystretch}{1}
\begin{equation} {\cal M}_{\tilde{e}} = \left( \begin{array} {cc}
        \me^{2} + \MMse^{2} - \frac{1}{2}\mZ^{2}\cos(2\tW)\cos(2\be)
                              & \me\left\{ \Ae + \mu\tan\be \right\} \\
        \me\left\{ \Ae + \mu\tan\be \right\} & \me^{2} + \Msle^{2}
                                          - \mZ^{2}\sin^{2}\tW\cos(2\be)
    \end{array} \right) \hspace*{1mm}. \end{equation}
The off--diagonal terms are proportional to the masses of the
corresponding non--\-super\-symmetric particles. Therefore, we will
assume that the mixing of the left-- and right scalar states are
negligible except in the stop and sbottom sector. The stop and sbottom
mass eigenstates we denote $\sba$, $\sbb$, $\sta$ and $\stb$. We
choose:
\[ \Msta>\Mstb \quad\quad \mbox{and}
     \quad\quad \Msba>\Msbb \hspace*{1mm}. \]
The masses and mixing angles ($\tht$ and $\tb$) can be found in
ref.~\cite{BEZ}.

\medskip

The couplings of the lightest CP--even Higgs boson to squarks and
sleptons without left--right mixing can be found in~\cite{GHKD}. The
couplings of the lightest CP--even Higgs boson to stops and sbottoms,
in the case with left--right scalar mixing, are given in
fig.~\ref{copl1f}. Only the couplings relevant for gluon fusion and
two--photon decay are listed. For special regions in parameter space,
the lower--generation squarks may give non--negligible contributions to
the one--loop decay width of the CP--even Higgses. We may also have
non--negligible contributions from sleptons.

\subsection{The chargino sector}

With $B$ the mass matrix of the charged Higgsino and gaugino
sector~\cite{Hab}, masses\myfoot{ Footnote on signs: In terms of
two--component Weyl spinors, the free part of the MSSM Lagrangian
density describing a Dirac particle is
\begin{equation} - i \left\{\xi\si^{m}\pad_{m}\oxi
     + \chi\si^{m}\pad_{m}\och \right\} + M\left\{ \chi\xi
        + \och\oxi \right\} \hspace*{1mm}. \label{HJhjjh} \end{equation}
The Dirac spinor can then be written
\[  \Psi = \left( \begin{array}{c}
                         \xi_{\al} \\
                         \och^{\alp}
                 \end{array} \right) \hspace*{1mm}. \]
The sign of the mass term is arbitrary. It can be changed by a
transformation
\[ (\xi,\oxi)\rightarrow (-\xi,-\oxi) \hspace*{1mm}, \]
with $\chi$ and $\och$ unchanged. In this paper we use the form chosen
in~(\ref{HJhjjh}). The sign of the mass term in the free part of the
Lagrangian density propagates to the equations of motion, to the
propagator and to the spin sum of $u\bar{u}$ and $v\bar{v}$. The sign
of the fermion mass factor in eq.~(\ref{HHGh}) below originates from
fermion propagators. The sign of the mass term propagates to couplings
involving fermions. In summary, the form, but not the value
of~(\ref{HHGh}), depends on the definition of the mass terms.}
and mass eigenstates $\chi_{1}^{\pm}$,  $\chi_{2}^{\pm}$ can be
calculated using
\jots5
\begin{eqnarray} \left( \begin{array}{c}
                  \chi_{1}^{+} \\
                  \chi_{2}^{+}
          \end{array} \right) &=& X\left( \begin{array}{c}
                                            -i\La^{+} \\
                                            \psi^{+}_{h}
                         \end{array} \right) \hspace*{1mm}, \nonumber \\
          \left( \begin{array}{c}
                  \chi_{1}^{-} \\
                  \chi_{2}^{-}
          \end{array} \right) &=& Y\left( \begin{array}{c}
                                            -i\La^{-} \\
                                            \psi^{-}_{H}
                       \end{array} \right) \hspace*{1mm}, \end{eqnarray}
where
\begin{eqnarray}
 X^{*}BY^{\ko}  &=&  \left( \begin{array}{cc}
                                   -\MC & 0 \\
                                   0 & -\mC
                  \end{array} \right) \quad ,
                           \quad\quad \MC \geq \mC \geq 0 \hspace*{1mm}.
     \label{HJHJjkjk} \end{eqnarray}
\jotr
The matrices $X$ and $Y$ must be unitary for the transformed free part
of the Lagrangian density to come out right. We assume that $X$ and $Y$
are real. Then the chargino masses will be real and the chargino sector
CP~invariant. We can choose
\jots5
\begin{eqnarray*} X & = & \left( \begin{array}{cc}
                \cos\tC & - \sin\tC \\
                \sin\tC & \cos\tC
              \end{array} \right) \hspace*{1mm}, \\*
    Y & = & \left( \begin{array}{cc}
                1 & 0 \\
                0 & \dC
             \end{array} \right)\left( \begin{array}{cc}
                                     \cos\pC & - \sin\pC \\
                                     \sin\pC & \cos\pC
                  \end{array} \right),
                     \quad\quad \dC=\pm 1 \hspace*{1mm}. \end{eqnarray*}
\jotr
Due to the parameter $\dC=\pm 1$, both masses can be chosen to have the
same sign.

We get the masses:
\jots{3}
\begin{eqnarray} \MC^{2} & = & S + \sqrt{T} \hspace*{1mm}, \nonumber \\*
    \mC^{2} & = & S - \sqrt{T} \hspace*{1mm}, \label{yutyTTy}
\end{eqnarray}
where
\begin{eqnarray*} S &=& \frac{1}{2}( \MNS^{2} + \mu^{2} )
                                            + \mW^{2} \hspace*{1mm}, \\
     T &=&\frac{1}{4}\left\{ \MNS^{2} - \mu^{2} \right\}^{2}
           + \mW^{4}\cos^{2}(2\be)
          + \mW^{2}\left\{ \MNS^{2} + \mu^{2} + 2\mu \MNS\sin(2\be)
                 \right\} \hspace*{1mm}.
\end{eqnarray*}
\jots{5}
The mixing angles $\tC$ and $\pC$ are given by:
\begin{eqnarray} \cos(2\tC) &=& \frac{\MNS^{2}-\mu^{2}
   +  2\mW^{2}\cos(2\be)}{\MC^{2}-\mC^{2}}\quad,
\quad\quad \frac{\tC}{ \MNS\sin\be + \mu\cos\be} \leq 0 \hspace*{1mm},
    \nonumber \\
    \cos(2\pC) &=& \frac{\MNS^{2}-\mu^{2}-2\mW^{2}\cos(2\be)}{\MC^{2}
    - \mC^{2}}\quad,
\quad\quad \frac{\pC}{ \MNS\cos\be + \mu\sin\be} \leq 0
\hspace*{1mm}. \label{chang} \end{eqnarray}
The signs of the masses are still not fixed. They may be reversed by
the addition of $\pi$ to one of the mixing angles,
\[ \pC \rightarrow \pC+\pi \Longrightarrow (\MC,\mC) \rightarrow
     (-\MC,-\mC) \hspace*{1mm}, \]
or
\[ \tC \rightarrow \tC+\pi \Longrightarrow (\MC,\mC) \rightarrow
    (-\MC,-\mC) \hspace*{1mm}. \]
We must choose a solution for $\tC$ and $\pC$ so that $-B$ is made
diagonal with non--negative entries, see eq.~(\ref{HJHJjkjk}).

\medskip

\renewcommand{\arraystretch}{1}
With $\mu=0$, as in the case with no left--right scalar mixing, the
lightest chargino tends to be rather light (see figs.~\ref{plssbc}
and~\ref{plfssbc}):
\begin{equation} \lim_{\tan\be\rightarrow\infty}\mC=0 \quad\quad
 \mbox{when} \quad\quad \mu=0 \hspace*{1mm}. \label{yuUHTTy}
\end{equation}

\medskip

The couplings of the lightest CP--even Higgs boson to charginos are
shown in fig.~\ref{copl2f}. Only the couplings relevant for two--photon
decay are listed.

\section{The decay widths}

\label{dewid}

In the CM system the one--loop contribution to the decay width for
$\hnl\rightarrow{g}{g}$ is given by~\cite{GHKD,CAN}
\begin{eqnarray} \Ga^{{g}{g}} &=& \frac{\as^{2}}{32\pi^{3}\mh}
\Biggl| - \sum_{k=1}^{2} C^{h}_{\stq_{k}}\left\{ 1 - \tau_{\stq_{k}}
    f(\tau_{\stq_{k}}) \right\}
  + 4\Mt C^{h}_{t}\left\{ 1 +\left[ 1 - \tau_{t}\right]f(\tau_{t})
    \right\}
  + \left( t,\widetilde{t}\rightarrow b,\widetilde{b} \right)
    \nonumber \\
& & \hspace*{20mm}\mbox{}- \sum_{\makebox[0mm]{$
    \st{\squ=\suq,\scq,\sdq,\ssq}$}}
    C^{hL}_{\squ}\left\{ 1 - \tauLq f(\tauLq) \right\}
  + \left(L\rightarrow R\right)  \Biggr|^{\Disp{2}} \hspace*{1mm}.
    \label{iUuiO}
\end{eqnarray}
Similarly, the one--loop contribution to the decay width for
$\hnl\rightarrow\ga\ga$ is given by
\jotr
\begin{eqnarray} \Ga^{\ga\ga} & = & \frac{\al^{2}}{64\pi^{3}\mh}
    \label{HHGh} \\*
& &
\times\Biiggls{6mm}| N\equ^{2}\left[ - \sum_{k=1}^{2}
      C^{h}_{\stq_{k}}\left\{ 1 - \tau_{\stq_{k}} f(\tau_{\stq_{k}})
      \right\}
    + 4\Mt C^{h}_{t}\left\{ 1 +\left[ 1 - \tau_{t}\right]f(\tau_{t})
      \right\} \right]
    + \left(t,\widetilde{t}\rightarrow b,\widetilde{b} \right)
      \nonumber \\[4mm]
& & \hspace*{17mm} \mbox{} - N \sum_{
    \makebox[0mm]{$\st{q=u,c,d,s}$}}\eq^{2} C^{hL}_{\squ}\left\{ 1 -
    \tauLq f(\tauLq) \right\}
    - \sum_{l=e,\mu,\tau} C^{hL}_{\alep}\left\{ 1 - \tauLl f(\tauLl)
      \right\}
    + \left(L\rightarrow R\right) \nonumber \\[4mm]
& & \hspace*{17mm} \mbox{} - C^{h}_{H}\left\{ 1 - \tau_{H} f(\tau_{H})
    \right\}
    + \frac{C^{h}_{W}}{\tau_{W}}\left\{ 2 + 3\tau_{W} + 3\tau_{W}
      \left[ 2-\tau_{W}\right]f(\tau_{W}) \right\} \nonumber \\*[4mm]
& & \hspace*{17mm} \mbox{} + 4\MC C^{h}_{\chi_{1}}\left\{ 1 +\left[ 1
    - \tau_{\chi_{1}}\right]f(\tau_{\chi_{1}}) \right\}
    + 4\mC C^{h}_{\chi_{2}}\left\{ 1 +\left[ 1 - \tau_{\chi_{2}}\right]
      f(\tau_{\chi_{2}}) \right\} \Biiggrso{6mm}|{\Disp{2}}
     \hspace*{1mm}, \nonumber \end{eqnarray}
where $N=3$ is the number of quark colours, and $\eq$ is the electric
charge of quark $q$. The couplings appearing in eqs.~(\ref{iUuiO})
and~(\ref{HHGh}) can be found in ref.~\cite{GHKD}, and
figs.~\ref{copl1f} and~\ref{copl2f}. In the case of no left--right
scalar mixing, there is no substantial simplification of these
expressions, only the couplings become a little simpler. The complex
function $f(\tau)$ is shown in fig.~\ref{plfunk}. It is given by
\renewcommand{\arraystretch}{1}
\begin{equation} f(\tau)= \frac{1}{2}\int_{0}^{1}
  \frac{\mbox{{\rm d}}x}{x}\ln\bigl[ 1-\frac{4x}{\tau}(1-x) \bigr]
= \left\{ \begin{array}{cl}\left[ \sin^{-1}(\sqrt{\frac{1}{\tau}})
  \right]^{2}  & \mbox{if}\quad \tau\geq 1 \hspace*{1mm}, \\
 - \left[ -\cosh^{-1}(\frac{1}{\sqrt{\tau}})+\frac{i\pi}{2}
   \right]^{2}  & \mbox{if}\quad \tau< 1 \hspace*{1mm}.
\end{array} \right. \label{hjjkdf} \end{equation}
The arguments $\tau_{i}$ appearing in eqs.~(\ref{iUuiO})
and~(\ref{HHGh}) are given by
\begin{equation} \tau_{i}=4\left(\frac{m_{i}}{\mh}\right)^{2}
      \hspace*{1mm}. \label{JKgfgfd} \end{equation}

\medskip

QCD corrections to the gluon--fusion cross--section have been
calculated in ref.~\cite{GSZ}. The QCD corrections to the two--photon
decay have been calculated in ref.~\cite{DK}. These corrections are not
included in our calculations.

\section{Numerical results}

\label{numd}

The effect of the left--right scalar mixing on the process
${g}{g}\rightarrow \hnl \rightarrow \ga\ga$ is illustrated by studying
the ratio $F$ of eq.~(\ref{OiuY}). The case of no left--right scalar
mixing is obtained by setting $\At=\Ab=\mu=0$, see eqs.~(\ref{uptypq})
and~(\ref{downtypq}).

\medskip

The decay rates, and hence the ratio $F$, will in general depend on a
large number of parameters, including the masses and coupling
constants. For the purpose of illustrating that the mixing of the
left-- and right--scalar states can have dramatic effects on the decay
rates, we shall consider only a rather constrained (five--dimensional)
parameter space. Therefore, we impose (see eq.~(\ref{JKhy}))
\[ \MsQ \equiv \MsQT = \MsT = \MsB \hspace*{1mm}, \]
\[ \MsQU = \Msu = \Msd = \MMse =
 \Msle \hspace*{1mm}, \]
\begin{equation} A \equiv \At = \Ab \hspace*{1mm}. \end{equation}
Further, we fix the parameters $\Mt$, $\MNS$ and $\MsQU$:
\begin{equation} \Mt=140~\GeV, \quad \MsQU=150~\GeV \quad \mbox{and}
       \quad \MNS=200~\GeV \hspace*{1mm}. \label{fixpar} \end{equation}
We have investigated the region in parameter space where
\begin{equation} \begin{array}{ccc}
    \tan\be \in [1.1,50], & \mA \in [20~\GeV,500~\GeV],
   & \MsQ \in [0,1000~\GeV], \\
    A \in [0,1000~\GeV], & \mu \in [-1000~\GeV,1000~\GeV] \hspace*{1mm}.
    \end{array} \label{hjkhjk} \end{equation}

\subsection{Mixing and masses}

The mixing between left-- and right scalars will show up in the ratio
$F$ as given by eq.~(\ref{OiuY}) partly through modified couplings, but
also through the effect mixing has on the masses. Obviously, the stop
and sbottom masses depend on the left--right scalar mixing, but also
the chargino and CP--even Higgs masses depend on the mixing. The
chargino masses depend on the mixing parameter $\mu$, see
eq.~(\ref{yutyTTy}), and the CP--even Higgs masses depend on the mixing
through radiative corrections.

\medskip

We can find regions with very high values for the ratio $F$. In these
regions, the effect of the mixing is mainly brought about by the change
in the masses. Either we have squarks close to threshold in the case of
left--right scalar mixing, or we have cancellations between diagrams in
the no--mixing case where the mass of the lightest chargino plays an
essential role. This will be discussed further in the following
sub--sections.

Corresponding to special regions in parameter space with high values
for the ratio $F$, we show in figs.~\ref{plssbc} to~\ref{pllhe} some of
the relevant masses. In figs.~\ref{plssbc} and~\ref{plfssbc}, we show
the influence of mixing on the masses of the lightest stop, sbottom and
chargino. In figs.~\ref{plsh}, \ref{plfsh} and~\ref{pllhe} we show the
influence of mixing on the mass of the lightest CP--even Higgs boson.

\subsection{Gross features of $F$}

It turns out that the ratio $F$, which gives the effects of the mixing
of left-- and right--scalar states on the production and decay of the
lightest CP--even Higgs boson, can take on values ranging from below
$10^{-6}$ to over $10^{4}$. However, such extreme values are not very
likely, they require very special combinations of parameters.

\medskip

In order to develop some intuition for how likely such extreme values
are, we show in fig.~\ref{figov} the distribution of $F$ in histogram
form, where each entry corresponds to one combination of the five
parameters $\tan\be$, $\mA$, $\MsQ$, $A$ and $\mu$. The plot is
produced by sampling $F$ over the five--dimensional parameter
space~(\ref{hjkhjk}), taking cells of equal size,
\begin{equation}
\Del\tan\be\times\Del\mA\times \Del \MsQ\times \Del A\times\Del\mu
 = 2\times 20~\GeV \times 20~\GeV \times 20~\GeV \times 20~\GeV
\hspace*{1mm}, \label{hjkul} \end{equation}
subject to the constraints
\jots{3}
\begin{eqnarray}
\Msq & > & \frac{\mZ}{2} \hspace*{1mm}, \nonumber \\
\Msq & > & \frac{\mh}{2} \hspace*{1mm}, \nonumber \\
 \mh & > &  30~\GeV \hspace*{1mm}, \nonumber \\
 \mC & > & 30~\GeV
\hspace*{1mm}. \label{hjRTEe} \end{eqnarray}
\jotr
The last condition only applies when $\left|\mu\right| \neq 0$. In
eq.~(\ref{hjRTEe}) $\Msq$ refers to the lightest squark or slepton. A
certain number of cells must be omitted since they yield unphysical
(complex) masses. Observe that we exclude regions where the lightest
CP--even Higgs boson can decay into pairs of squarks or sleptons. The
experimental bounds on the squark and Higgs masses are parameter
dependent. In ref.~\cite{DLAO} results from all the LEP experiments are
included to give mass bounds for the $h^{0}$. Almost the entire
accessible mass region at LEP-I is excluded ($\mh>40~\GeV$), but mixing
is neglected.

It should be noted that $F$ is sharply peaked at $F=1$, i.e. in most of
parameter space the effect of the mixing is rather small. However,
there are small but finite regions in parameter space where the ratio
$F$ deviates substantially from unity. These will be explored in the
following sub--sections.

\subsection{Regions of limited corrections}

We have seen in fig.~\ref{figov} that the corrections to
$\Ga^{{g}{g}}\Ga^{\ga\ga}$ due to mixing between the left-- and right
scalars can take on rather large values in various parts of parameter
space. The question then arises: Are there ``safe'' regions in some
sub--space of the five--dimensional parameter space~(\ref{hjkhjk})
where these corrections are ``small''? Basically, the answer is no.
This is illustrated in fig.~\ref{plli}, where we indicate in black
regions where $0.1<F<10$ in (a) the $\mu$--$A$ plane, and (b) the
$\tan\be$--$\MsQ$ plane, for {\em any} values of the other three
parameters within the domain~(\ref{hjkhjk}). In our search for ``safe''
regions, the parameters $\MsQ$, $\mu$ and $A$ have been scanned
according to the step sizes given in table~\ref{stpsiz}.

\subsection{Variation of $F$ with $\tan\be$ and $\mA$}

There are many regions in the five--dimensional parameter
space~(\ref{hjkhjk}) where the ratio $F$ differs substantially from
unity. Scanning parameter space, we find two effects causing ``very''
high values for $F$. Either we have enhancements of the decay rates in
the case of left--right scalar mixing due to squarks being close to
threshold, or there is destructive interference in the no--mixing case.
In the regions with rather ``extreme'' high values of $F$ due to
destructive interference, the lightest chargino plays the role of
cancelling the imaginary part of the contribution from the bottom
quark, see eq.~(\ref{HHGh}).

In order to locate some of these regions, we show in figs.~\ref{pla}
and~\ref{plfa} contour plots of $F$ in the $\tan\be$--$\mA$ plane. The
parameters $\Mt$, $\MsQU$ and $\MNS$ are taken at their default
values~(\ref{fixpar}). For the remaining three parameters we consider
two sets of values, $\MsQ=269~\GeV$, $A=517~\GeV$ and $\mu=775~\GeV$ in
fig.~\ref{pla}, and $\MsQ=303~\GeV$, $A=32~\GeV$ and $\mu=821~\GeV$ in
fig.~\ref{plfa}.

\medskip

For the first set of parameters, the ratio $F$ has a rather dramatic
variation, from less than 0.05 at high $\mA$ and ``medium'' values of
$\tan\be$, to values above $10^{3}$ at $\tan\be\approx 7$ and
$\mA\approx 150~\GeV$, as illustrated in fig.~\ref{pla}. This maximum
value of $F$ actually occurs within the ``problematic'' region where
the no--mixing calculation yields a $\ga\ga$ decay rate that is too low
for discovery of the $\hnl$, cf. fig.~30 in ref.~\cite{ZKFZ}. The
maximum value of $F$ shown in fig.~\ref{pla} ($F_{\mbox{{ \footnotesize
max}}}\approx 5.6\cdot 10^{3}$) is due to enhancement mostly of
$\Ga^{{g}{g}}_{\mbox{{ \footnotesize mixing}}}$ but also of
$\Ga^{\ga\ga}_{\mbox{{ \footnotesize mixing}}}$. This enhancement is
due to the contributions of the lightest stop to the
expressions~(\ref{iUuiO}) and~(\ref{HHGh}). In the  region of parameter
space where this maximum occurs, the lightest stop has a mass close to
threshold in the case of left--right scalar mixing, but far above
threshold in the no--mixing case. Thus, in the case of mixing, the
variable $\tau_{\stq_{2}} \raisebox{-2mm}{\Dis{\stackrel{{\Disp
>}}{{\Disp \sim}}}} 1$. Without mixing, $A=\mu=0$, $\tau_{\stq_{2}} \gg
1$ and the corresponding $f$ becomes small. The mass of the lightest
stop is in the relevant region shown in fig.~\ref{plssbc}.

\medskip

For the second set of parameters ($\MsQ=303~\GeV$, $A=32~\GeV$ and
$\mu=821~\GeV$) there is an even more dramatic variation of $F$, from
less than 0.2 to values above $10^{11}$, as illustrated in
fig.~\ref{plfa}. This maximum is due to $\Ga^{\ga\ga}_{\mbox{{
\footnotesize no mixing}}}$, which becomes very small because of
destructive interference between the contributing diagrams.
Consequently, the ratio $F$ becomes anomalously large ($F_{\mbox{{
\footnotesize max}}}\approx 5\cdot 10^{11}$). The bottom quark and the
lightest chargino (we ignore the leptons and lower--generation quarks)
are in the no--mixing case the only particles below threshold, and they
therefore give the only imaginary contributions to the decay
amplitudes. Their imaginary contributions cancel at this maximum of
$F$. However, this case is close to being unphysical, since with no
left--right scalar mixing, the lightest chargino tends to be too light
(see eq.~(\ref{yuUHTTy}) and figs.~\ref{plssbc} and~\ref{plfssbc}).

\medskip

Scanning parameter space, all the ``extreme'' maxima of $F$  we find
are due to the two effects mentioned above. Either squarks are close to
threshold (as in fig.~\ref{pla}), or there are cancellations where the
lightest chargino plays an essential role (as in fig.~\ref{plfa}). But
even when these conditions are not met, the effects of the mixing may
still be significant. This is illustrated in fig.~\ref{fighv} where we
ignore the chargino sector (by putting the chargino--Higgs couplings
equal to zero). Furthermore, we impose the constraint that all squark
and slepton masses exceed $80~\GeV$. Thus, except for leptons and
quarks, all the contributing particles are far above threshold. By
ignoring the chargino sector, we get rid of the (unphysical?)
cancellations which can make the ratio $F$ anomalously large. In
fig.~\ref{fighv} we show a surface plot of the ratio
$F=(\Ga^{{g}{g}}\Ga^{\ga\ga})_{\mbox{{ \footnotesize mixing}}}/
        (\Ga^{{g}{g}}\Ga^{\ga\ga})_{\mbox{{ \footnotesize no mixing}}}$
in the $\tan\be$--$\mA$ plane, with $\Mt$, $\MsQU$ and $\MNS$ given by
eq.~(\ref{fixpar}). Furthermore, for each point in the $\tan\be$--$\mA$
plane, the values chosen for the parameters $\MsQ$, $A$ and $\mu$ are
those which maximize the ratio $F$.

\section{Conclusions}

\label{conc}

In most of parameter space, the effect of the left--right scalar mixing
on the process ${g}{g}\rightarrow \hnl \rightarrow \ga\ga$ is small.
However, there exist regions where the mixing is important. According
to fig.~\ref{figov}, the left--right scalar mixing can either suppress
or enhance the cross sections relative to the no--mixing approximation.
In this study, we have been mostly concerned with regions where we find
large enhancements of the cross sections when the mixing is
incorporated. A more comprehensive study will be presented elsewhere.

When the lightest stop and/or sbottom is close to (or below) threshold
when left--right scalar mixing is considered, but far above threshold
in the no--mixing approximation, mixing can lead to large values of the
ratio
\[ F = \frac{\Ga^{{g}{g}}_{\mbox{{ \footnotesize
mixing}}}}{\Ga^{{g}{g}}_{\mbox{{ \footnotesize no mixing}}}}
        \times \frac{\Ga^{\ga\ga}_{\mbox{{ \footnotesize
mixing}}}}{\Ga^{\ga\ga}_{\mbox{{ \footnotesize no mixing}}}}
        \hspace*{1mm}. \]
We find such large values of $F$ within the ``problematic'' region
where the no--mixing calculation yields a $\ga\ga$ decay rate that is
too low for discovery of the $\hnl$, cf. fig.~30 in ref.~\cite{ZKFZ}.

Mixing can also introduce large corrections in regions where, due to a
high degree of cancellations between diagrams, the no--mixing cross
section nearly vanishes. The ratio $F$ can become ``anomalously''
large, but only when the contribution from the lightest chargino is
essential. However, in most of parameter space the no--mixing
approximation is somewhat unphysical, since it tends to make the
lightest chargino too light.

Corrections due to mixing will also show up through the
mixing--dependent couplings. These effects are much more moderate, but
still quite significant, as shown in fig.~\ref{fighv}.

\section*{Acknowledgement}

It is a pleasure to thank the CERN Theory Division for kind hospitality
during a stay as a visiting student. I especially wish to thank John
Ellis and Fabio Zwirner for valuable suggestions and support. I am
indebted to Per Osland for checking my calculations and correcting my
English spelling and style. Per Steinar Iversen has been very helpful
with computer questions. This research has been supported by the
Research Council of Norway.

\clearpage

\section*{Table caption}

\setcounter{table}{0}
\begin{description}
\stepcounter{table}
\item[Table~\thetable:] Fields and other symbols.
\stepcounter{table}
\item[Table~\thetable:] Step sizes employed for figure~\ref{plli}.
\end{description}
\setcounter{table}{0}

\clearpage

\small
\renewcommand{\baselinestretch}{1.6}
\normalsize
\renewcommand{\arraystretch}{1}
\setlength{\tabcolsep}{\rtcol}
\newsavebox{\tabl}
\sbox{\tabl}
{\begin{tabular}{||c|l||}\hline\hline
$w^{\alpha}$  & U(1) supersymmetric field strength, see~\cite{MuKi}
or~\cite{BaWe} \\[0.3mm]
$W^{\alpha}$  & SU(2) supersymmetric field strength \\[0.3mm]
$W_{S}^{\alpha}$  & SU(3) supersymmetric field strength \\[0.3mm] \hline
$v$       & U(1) vector superfield, see~\cite{MuKi} or~\cite{BaWe}
\\[0.3mm]
$V$                               & SU(2) vector superfields, $V=\sum
T^{K}_{\mbox{\scriptsize SU(2)}}V^{K}$ \\[1.5mm]
$V_{S}$                           & SU(3) vector superfields,
                                                    $V_{S}=\sum
T^{K}_{\mbox{\scriptsize SU(3)}}V^{K}_{S}$ \\[0.3mm]
$\what{3}{0.7}{E}^{R}$  &     U(1) chiral superfield for leptons,
see~\cite{MuKi} or~\cite{BaWe} \\[0.3mm]
$\what{3}{0.7}{D}^{R}$  &     U(1) chiral superfield for d--type
quarks \\[0.3mm]
$\what{3}{0.2}{U}^{R}$  &     U(1) chiral superfield for u--type
quarks \\[0.3mm]
$\what{3}{0.7}{L}$      &     SU(2) doublet chiral superfield for
leptons \\[0.3mm]
$\what{3}{0.7}{Q}$      &     SU(2) doublet chiral superfield for
quarks \\[0.3mm]
$\what{3}{0.7}{H}_{1}$                        &      $Y=-\frac{1}{2}$
SU(2) doublet chiral Higgs superfield \\[0.3mm]
$\what{3}{0.7}{H}_{2}$                        &      $Y=\frac{1}{2}$
SU(2) doublet chiral Higgs superfield \\[0.3mm]
$\what{3}{0.7}{Q}^{L}_{\mbox{\scriptsize SU(3)}}$   &      SU(3)
triplet chiral superfield for left--chiral quarks \\[1mm]
$\what{3}{0.7}{Q}^{R}_{\mbox{\scriptsize SU(3)}}$   &      SU(3)
triplet chiral superfield for right--chiral quarks \\[1mm] \hline
$\tilde{{e}}^{R}$        \ned{4}             &      Scalar part of the
U(1) electron--type chiral lepton superfield \\[0.3mm]
$\tilde{{d}}_{1}^{R}$                               &      Scalar part
of the U(1) d--type chiral quark superfield \\[0.3mm]
$\tilde{{u}}^{R}$                               &      Scalar part of
the U(1) u--type chiral quark superfield \\[0.3mm]
$L$                                      &      Scalar part of the
SU(2) doublet chiral lepton superfield \\[0.3mm]
$Q$                                      &      Scalar part of the
SU(2) doublet chiral quark superfield \\[0.3mm]
$H_{1}$                                  &      Scalar part of the
$Y=-\frac{1}{2}$ SU(2) doublet chiral Higgs superfield \\[0.3mm]
$H_{2}$                                  &      Scalar part of the
$Y=\frac{1}{2}$ SU(2) doublet chiral Higgs superfield \\[0.3mm]
$\la$                                    &      U(1) gaugino field
\\[0.3mm]
$\Lambda_{I}$                                &      SU(2) gaugino
fields \\[0.3mm]
\hline\hline
\end{tabular}}

\begin{table}

\begin{center}
\setlength{\unitlength}{1cm}
\begin{picture}(0,17)
\put(0,6.5){\makebox(0,16)[b]{\usebox{\tabl}}}
\end{picture}

\refstepcounter{table}
\label{kjtyFG}

\vspace*{38mm}

Table~\thetable

\end{center}
\end{table}

\small
\renewcommand{\baselinestretch}{2.5}
\normalsize

\clearpage

\sbox{\Forh}{$\wtilde{3}{0.8}{M}_{\hspace*{-1.2mm}E}$,
$\wtilde{3}{0.2}{m}_{\!E}$, $\wtilde{3}{0.8}{M}_{\hspace*{-1mm}U}$,
$\wtilde{3}{0.2}{m}_{U}$,}
\sbox{\Abox}{$\wtilde{3}{0.2}{m}_{\!D}$, $M_{\!H_{1}}$, $M_{\!H_{2}}$,
$m_{{\displaystyle \tilde{{\scriptstyle \lambda}}}}$,
$m_{{\displaystyle \tilde{{\scriptstyle \Lambda}}}}$}
\begin{table}
\begin{center}

\begin{tabular}{||c|l||}\hline\hline
$m_{{e}}$, $m_{{d}}$, $m_{{u}}$ \ned{4} & Charged lepton, d--type and
u--type quark masses \\[5mm]
\begin{minipage}{4.398541503cm}
\small
\renewcommand{\baselinestretch}{1.3}
\normalsize
\begin{center}
\usebox{\Forh} \\ \usebox{\Abox}
\end{center}
\end{minipage} & Soft--breaking mass parameters \\[5mm]
$A_{{e}}$, $A_{{d}}$, $A_{{u}}$ & Soft--breaking Yukawa parameters
 \\[0.4mm] \hline
$g_{S}$  \ned{4}         &      SU(3) charge \\[0.3mm]
$g$                      &      SU(2) charge \\[0.3mm]
$g'$                     &      U(1) charge \\[7mm]
$\tan\be$                &  \begin{minipage}{10.61737371cm}
\small
\renewcommand{\baselinestretch}{1.3}
\normalsize
Ratio of the vacuum expectation values of the Higgs fields,
$\tan\be=v_{2}/v_{1}$. The vacuum expectation value $v_{i}$ belongs to
the scalar SU(2) doublet $H_{i}$.
\end{minipage} \\[7mm]
$\beta^{Hh}$ &  Soft Higgs--Higgs coupling parameter \\[0.3mm]
$\epsilon$   &  The matrix given by: $\epsilon_{11}=\epsilon_{22}=0$
and $\epsilon_{12}=-\epsilon_{21}=1$ \\[0.2mm]
\hline\hline
\end{tabular}

\vspace*{10mm}

Table~\thetable~(continued)

\end{center}
\end{table}

\small
\renewcommand{\baselinestretch}{2}
\normalsize

\clearpage

\small
\renewcommand{\baselinestretch}{2.5}
\normalsize
\renewcommand{\arraystretch}{1}
\setlength{\tabcolsep}{\rtcol}
\begin{table}
\begin{center}
\begin{tabular}{||c|c|c||}\hline\hline
Parameter & Interval & Step sizes \\ \hline
$\MsQ$, $\left|\mu\right|$ , $A$ $[\GeV]$ & $\left[0,10\right]$,
$\left(10,100\right]$, $\left(100,1000\right]$ & 1, 10, 50 $\GeV$
\\ \hline
$\mA$ $[\GeV]$ & $\left[0,40\right]$, $\left(40,100\right]$,
$\left(100,500\right]$ & 2, 10, 40 $\GeV$ \\ \hline
$\tan\be$ & $\left[1.1,10\right]$, $\left(10,50\right]$ & 1, 5
\\ \hline \hline
\end{tabular}
\refstepcounter{table}
\label{stpsiz}

\vspace*{10mm}

Table~\thetable

\end{center}
\end{table}

\small
\renewcommand{\baselinestretch}{2}
\normalsize

\clearpage

\section*{Figure captions}

\small
\renewcommand{\baselinestretch}{2}
\normalsize
\setcounter{figure}{0}
\begin{description}
\stepcounter{figure}
\item[Figure~\thefigure:] Couplings of the lightest CP--even Higgs
boson to stops and sbottoms in the case with left--right scalar mixing.
Only the couplings relevant for gluon fusion and two--photon decay are
listed.
\stepcounter{figure}
\item[Figure~\thefigure:] Couplings of the lightest CP--even Higgs
boson to charginos. Only the couplings relevant for two--photon decay
are listed.
\stepcounter{figure}
\item[Figure~\thefigure:] The function $f(\tau)$ of
equation~(\ref{hjjkdf}).
\stepcounter{figure}
\item[Figure~\thefigure:] Masses of the lightest stop and sbottom
squarks, as well as the lightest chargino mass, as functions of
$\tan\be$, for the parameters of figs.~\ref{pla} and~\ref{plsh}, i.e.
for $\Mt$, $\wtilde{3}{0.8}{M}_{\hspace*{-1mm}U}$ and
$m_{{\displaystyle \tilde{{\scriptstyle \Lambda}}}}$ given by
eq.~(\ref{fixpar}), and for $\MsQ=269~\GeV$, $A=517~\GeV$ and
$\mu=775~\GeV$. To lowest order, the chargino mass and squark masses do
not depend on the mass $\mA$ of the CP--odd Higgs boson.
\stepcounter{figure}
\item[Figure~\thefigure:] Masses of the lightest stop and sbottom
squarks, as well as the lightest chargino mass, as functions of
$\tan\be$, for the parameters of figs.~\ref{plfa} and~\ref{plfsh}, i.e.
for $\Mt$, $\wtilde{3}{0.8}{M}_{\hspace*{-1mm}U}$ and
$m_{{\displaystyle \tilde{{\scriptstyle \Lambda}}}}$ given by
eq.~(\ref{fixpar}), and for $\MsQ=303~\GeV$, $A=32~\GeV$ and
$\mu=821~\GeV$.
\stepcounter{figure}
\item[Figure~\thefigure:] Mass of the lightest CP--even Higgs boson for
the parameters of fig.~\ref{pla}, i.e. for $\Mt$,
$\wtilde{3}{0.8}{M}_{\hspace*{-1mm}U}$ and $m_{{\displaystyle
\tilde{{\scriptstyle \Lambda}}}}$ given by eq.~(\ref{fixpar}), and for
$\MsQ=269~\GeV$, $A=517~\GeV$ and $\mu=775~\GeV$.
\stepcounter{figure}
\item[Figure~\thefigure:] Mass of the lightest CP--even Higgs boson for
the parameters of fig.~\ref{plfa}, i.e. for $\Mt$,
$\wtilde{3}{0.8}{M}_{\hspace*{-1mm}U}$ and $m_{{\displaystyle
\tilde{{\scriptstyle \Lambda}}}}$ given by eq.~(\ref{fixpar}), and for
$\MsQ=303~\GeV$, $A=32~\GeV$ and $\mu=821~\GeV$.
\stepcounter{figure}
\item[Figure~\thefigure:] Surface plot showing the mass of the lightest
CP--even Higgs boson for the parameters of fig.~\ref{fighv}, i.e. for
$\Mt$, $\wtilde{3}{0.8}{M}_{\hspace*{-1mm}U}$ and $m_{{\displaystyle
\tilde{{\scriptstyle \Lambda}}}}$ given by eq.~(\ref{fixpar}). For each
point in the $\tan\be$--$\mA$ plane, the values chosen for the
parameters $\MsQ$, $A$ and $\mu$ are those which maximize the ratio
$F=(\Ga^{{g}{g}}\Ga^{\gamma\gamma})_{\mbox{{\footnotesize
mixing}}}/(\Ga^{{g}{g}}\Ga^{\gamma\gamma})_{\mbox{{\footnotesize no
mixing}}}$. When calculating this ratio $F$, we neglect the
contributions from the charginos. Furthermore, we impose the constraint
that all squark and slepton masses exceed $80~\GeV$.
\stepcounter{figure}
\item[Figure~\thefigure:] Number of cells for different values of
$F=(\Ga^{{g}{g}}\Ga^{\gamma\gamma})_{\mbox{{\footnotesize
mixing}}}/(\Ga^{{g}{g}}\Ga^{\gamma\gamma})_{\mbox{{\footnotesize no
mixing}}}$. The parameter space (\ref{hjkhjk}) is sampled according to
equal--size cells given by eq.(\ref{hjkul}), and $\Mt$,
$\wtilde{3}{0.8}{M}_{\hspace*{-1mm}U}$ and $m_{{\displaystyle
\tilde{{\scriptstyle \Lambda}}}}$ are taken at their default
values~(\ref{fixpar}).
\stepcounter{figure}
\item[Figure~\thefigure:] Regions in parameter space where the
corrections due to the mixing of left-- and right scalar states are
limited to within a factor 10. In black: $0.1<F<10$ in (a) the
$\mu$--$A$ plane, and (b) the $\tan\be$--$\MsQ$ plane
\stepcounter{figure}
\item[Figure~\thefigure:] Contour plots of the ratio
$F=(\Ga^{{g}{g}}\Ga^{\gamma\gamma})_{\mbox{{\footnotesize
mixing}}}/(\Ga^{{g}{g}}\Ga^{\gamma\gamma})_{\mbox{{\footnotesize no
mixing}}}$ in the $\tan\be$--$\mA$ plane, with $\Mt$,
$\wtilde{3}{0.8}{M}_{\hspace*{-1mm}U}$ and $m_{{\displaystyle
\tilde{{\scriptstyle \Lambda}}}}$ given by eq.~(\ref{fixpar}).
Furthermore, we have here chosen $\MsQ=269~\GeV$, $A=517~\GeV$ and
$\mu=775~\GeV$. The area labelled ``Forbidden region'' is ruled out by
the constraint~(\ref{hjRTEe}), or by the mass of the the lightest
CP--even Higgs boson being complex.
\stepcounter{figure}
\item[Figure~\thefigure:] Contour plots of the ratio
$F=(\Ga^{{g}{g}}\Ga^{\gamma\gamma})_{\mbox{{\footnotesize
mixing}}}/(\Ga^{{g}{g}}\Ga^{\gamma\gamma})_{\mbox{{\footnotesize no
mixing}}}$ in the $\tan\be$--$\mA$ plane, with $\Mt$,
$\wtilde{3}{0.8}{M}_{\hspace*{-1mm}U}$ and $m_{{\displaystyle
\tilde{{\scriptstyle \Lambda}}}}$ given by eq.~(\ref{fixpar}).
Furthermore, we have here chosen $\MsQ=303~\GeV$, $A=32~\GeV$ and
$\mu=821~\GeV$. The areas labelled ``Unphysical region'' are ruled out
by the mass of the CP--even Higgs boson being complex. In (a) a larger
area of the $\tan\be$--$\mA$ plane is shown, whereas (b) is devoted to
the maximum.
\stepcounter{figure}
\item[Figure~\thefigure:] Surface plot of the ratio
$F=(\Ga^{{g}{g}}\Ga^{\gamma\gamma})_{\mbox{{\footnotesize
mixing}}}/(\Ga^{{g}{g}}\Ga^{\gamma\gamma})_{\mbox{{\footnotesize no
mixing}}}$ in the $\tan\be$--$\mA$ plane, with $\Mt$,
$\wtilde{3}{0.8}{M}_{\hspace*{-1mm}U}$ and $m_{{\displaystyle
\tilde{{\scriptstyle \Lambda}}}}$ given by eq.~(\ref{fixpar}). For each
point in the $\tan\be$--$\mA$ plane, the values chosen for the
parameters $\MsQ$, $A$ and $\mu$ are those which maximize the ratio
$F$. We here exclude ``extreme'' (unphysical?) contributions by
neglecting the couplings to charginos. Furthermore, we impose the
constraint that all squark and slepton masses exceed $80~\GeV$.
\end{description}
\setcounter{figure}{0}

\clearpage

\newlength{\lla}
\setlength{\lla}{14.77129948cm}
\newfont{\feyn}{bolge}
\newsavebox{\fsymb}

\begin{figure}
\refstepcounter{figure}
\label{copl1f}

\begin{center}
\settowidth{\unitlength}{\feyn E}
\sbox{\fsymb}{
\parbox{\unitlength}{\begin{picture}(1,0.7)(0.1,0.15)
\put(0,0.5){{\feyn E}}
\put(0.1,0.53){\makebox(0,0)[lb]{{\normalsize $h^{0}$}}}
\put(0.88,0.63){\makebox(0,0)[lb]{{\normalsize $\sta$}}}
\put(0.88,0.35){\makebox(0,0)[lt]{{\normalsize $\sta$}}}
\end{picture}}}
\setlength{\unitlength}{1cm}
\[ \rule[-13mm]{0mm}{26mm}\begin{picture}(14,0)(0.7,0)
\put(0,0){\usebox{\fsymb}}
 \put(2,0){\makebox(0,0)[l]{\begin{minipage}{\lla}
 \vspace{-\abovedisplayskip}
 \begin{eqnarray*}
C^{h}_{\tilde{t}_{1}}&=&  \frac{ig\mZ}{2\cos\tW}\sin(\alpha+\be)
  \left[ - \frac{4}{3}\sin^{2}\tW\cos(2\tht)+\cos^{2}\tht\right]
   \nonumber \\
 & & \mbox{}+ \frac{ig\Mt}{2m_{\ssst{{ W}}}\sin\beta}\left[
     \left(\mu\sin\alpha-\At\cos\alpha\right)\sin(2\tht)
    - 2\Mt\cos\alpha\right]
 \end{eqnarray*}\end{minipage}}}
\end{picture} \]
\settowidth{\unitlength}{\feyn E}
\sbox{\fsymb}{
\parbox{\unitlength}{\begin{picture}(1,0.7)(0.1,0.15)
\put(0,0.5){{\feyn E}}
\put(0.1,0.53){\makebox(0,0)[lb]{{\normalsize $h^{0}$}}}
\put(0.88,0.63){\makebox(0,0)[lb]{{\normalsize $\stb$}}}
\put(0.88,0.35){\makebox(0,0)[lt]{{\normalsize $\stb$}}}
\end{picture}}}
\setlength{\unitlength}{1cm}
\[ \rule[-13mm]{0mm}{26mm}\begin{picture}(14,0)(0.7,0)
\put(0,0){\usebox{\fsymb}}
 \put(2,0){\makebox(0,0)[l]{\begin{minipage}{\lla}
 \vspace{-\abovedisplayskip}
 \begin{eqnarray*}
C^{h}_{\tilde{t}_{2}}&=& \frac{ig\mZ}{2\cos\tW}\sin(\alpha+\be)
  \left[ \frac{4}{3}\sin^{2}\tW\cos(2\tht)+\sin^{2}\tht\right]
   \nonumber \\
 & & \mbox{}- \frac{ig\Mt}{2m_{\ssst{{ W}}}\sin\beta}\left[
    \left(\mu\sin\alpha-\At\cos\alpha\right)\sin(2\tht)
    + 2\Mt\cos\alpha\right]
 \end{eqnarray*}\end{minipage}}}
\end{picture} \]
\settowidth{\unitlength}{\feyn E}
\sbox{\fsymb}{
\parbox{\unitlength}{\begin{picture}(1,0.7)(0.1,0.15)
\put(0,0.5){{\feyn E}}
\put(0.1,0.53){\makebox(0,0)[lb]{{\normalsize $h^{0}$}}}
\put(0.88,0.60){\makebox(0,0)[lb]{{\normalsize $\sba$}}}
\put(0.88,0.35){\makebox(0,0)[lt]{{\normalsize $\sba$}}}
\end{picture}}}
\setlength{\unitlength}{1cm}
\[ \rule[-13mm]{0mm}{26mm}\begin{picture}(14,0)(0.7,0)
\put(0,0){\usebox{\fsymb}}
 \put(2,0){\makebox(0,0)[l]{\begin{minipage}{\lla}
 \vspace{-\abovedisplayskip}
 \begin{eqnarray*}
C^{h}_{\tilde{b}_{1}}&=& \mbox{}- \frac{ig\mZ}{2\cos\tW}\sin(\alpha+\be)
    \left[ -\frac{2}{3}\sin^{2}\tW\cos(2\tb)+\cos^{2}\tb\right]
     \nonumber \\
 & & \mbox{}+ \frac{ig\Mb}{2m_{\ssst{{ W}}}\cos\beta}\left[
    - \left(\mu\cos\alpha-\Ab\sin\alpha\right)\sin(2\tb)
    + 2\Mb\sin\alpha\right]
 \end{eqnarray*}\end{minipage}}}
\end{picture} \]
\settowidth{\unitlength}{\feyn E}
\sbox{\fsymb}{
\parbox{\unitlength}{\begin{picture}(1,0.7)(0.1,0.15)
\put(0,0.5){{\feyn E}}
\put(0.1,0.53){\makebox(0,0)[lb]{{\normalsize $h^{0}$}}}
\put(0.88,0.60){\makebox(0,0)[lb]{{\normalsize $\sbb$}}}
\put(0.88,0.35){\makebox(0,0)[lt]{{\normalsize $\sbb$}}}
\end{picture}}}
\setlength{\unitlength}{1cm}
\[ \rule[-13mm]{0mm}{26mm}\begin{picture}(14,0)(0.7,0)
\put(0,0){\usebox{\fsymb}}
 \put(2,0){\makebox(0,0)[l]{\begin{minipage}{\lla}
 \vspace{-\abovedisplayskip}
 \begin{eqnarray*}
C^{h}_{\tilde{b}_{2}}&=& \frac{ig\mZ}{2\cos\tW}\sin(\alpha+\be)
    \left[ -\frac{2}{3}\sin^{2}\tW\cos(2\tb)-\sin^{2}\tb\right]
    \nonumber \\
 & & \mbox{} + \frac{ig\Mb}{2m_{\ssst{{ W}}}\cos\beta}\left[
      \left(\mu\cos\alpha-\Ab\sin\alpha\right)\sin(2\tb)
    + 2\Mb\sin\alpha\right]
 \end{eqnarray*}\end{minipage}}}
\end{picture} \]

\vspace*{10mm}

Figure~\thefigure
\end{center}
\end{figure}

\clearpage

\begin{figure}
\refstepcounter{figure}
\label{copl2f}

\begin{center}
\setlength{\lla}{14.77129948cm}
\settowidth{\unitlength}{\feyn F}
\sbox{\fsymb}{
\parbox{\unitlength}{\begin{picture}(1,0.7)(0.1,0.15)
\put(0,0.5){{\feyn F}}
\put(0.1,0.53){\makebox(0,0)[lb]{{\normalsize $h^{0}$}}}
\put(0.88,0.68){\makebox(0,0)[lb]{{\normalsize $\chi_{1}$}}}
\put(0.88,0.35){\makebox(0,0)[lt]{{\normalsize $\chi_{1}$}}}
\end{picture}}}
\setlength{\unitlength}{1cm}
\[ \rule[-13mm]{0mm}{26mm}\begin{picture}(14,0)(0.7,0)
\put(0,0){\usebox{\fsymb}}
 \put(2,0){\makebox(0,0)[l]{\begin{minipage}{\lla}
\centering
\Dis{ C^{h}_{\chi_{1}} =
    \frac{ig}{\sqrt{2}}\left[ \cos\alpha\sin\tC\cos\pC
   - \sin\alpha\cos\tC\sin\pC \right] }\end{minipage}}}
\end{picture} \]
\settowidth{\unitlength}{\feyn F}
\sbox{\fsymb}{
\parbox{\unitlength}{\begin{picture}(1,0.7)(0.1,0.15)
\put(0,0.5){{\feyn F}}
\put(0.1,0.53){\makebox(0,0)[lb]{{\normalsize $h^{0}$}}}
\put(0.88,0.68){\makebox(0,0)[lb]{{\normalsize $\chi_{2}$}}}
\put(0.88,0.35){\makebox(0,0)[lt]{{\normalsize $\chi_{2}$}}}
\end{picture}}}
\setlength{\unitlength}{1cm}
\[ \rule[-13mm]{0mm}{26mm}\begin{picture}(14,0)(0.7,0)
\put(0,0){\usebox{\fsymb}}
 \put(2,0){\makebox(0,0)[l]{\begin{minipage}{\lla}
\centering
\Dis{ C^{h}_{\chi_{2}} = \frac{ig\dC}{\sqrt{2}}\left[
  \sin\alpha\sin\tC\cos\pC - \cos\alpha\cos\tC\sin\pC \right] }
  \end{minipage}}}
\end{picture} \]

\vspace*{10mm}

Figure~\thefigure
\end{center}
\end{figure}

\clearpage

\begin{figure}
\refstepcounter{figure}
\label{plfunk}
\begin{center}
\mbox{\epsffile{ffunk.eps}}

\vspace{10mm}

Figure~\thefigure
\end{center}
\end{figure}

\clearpage

\begin{figure}
\refstepcounter{figure}
\label{plssbc}
\begin{center}
\setlength{\unitlength}{1cm}
\begin{picture}(16,0)
\put(3.7,-2){\makebox(0,0)[bl]{{\Large $\Mstb$ without mixing}}}
\put(4.3,-15.3){\makebox(0,0)[bl]{{\Large $\Mstb$ with mixing}}}
\put(3.7,-4.25){\makebox(0,0)[bl]{{\Large $\Msbb$ without mixing}}}
\put(7.5,-8.7){\makebox(0,0)[bl]{{\Large $\Msbb$ with mixing}}}
\put(6.4,-17){\makebox(0,0)[bl]{{\Large $\mC$ without mixing}}}
\put(8,-7.2){\makebox(0,0)[bl]{{\Large $\mC$ with mixing}}}
\end{picture}
\mbox{\epsffile{fig7m.eps}}

\vspace*{5mm}

Figure~\thefigure
\end{center}
\end{figure}

\clearpage

\begin{figure}
\refstepcounter{figure}
\label{plfssbc}
\setlength{\unitlength}{1cm}
\begin{center}
\begin{picture}(16,0)
\put(7.78,-2.16){\makebox(0,0)[bl]{{\Large $\Mstb$ without mixing}}}
\put(2.2,-7.6){\makebox(0,0)[bl]{{\Large $\Mstb$ with mixing}}}
\put(8.1,-4.1){\makebox(0,0)[bl]{{\Large $\Msbb$ without mixing}}}
\put(5.75,-6.3){\makebox(0,0)[bl]{{\Large $\Msbb$ with mixing}}}
\put(6.9,-17.05){\makebox(0,0)[bl]{{\Large $\mC$ without mixing}}}
\put(3.6,-9.05){\makebox(0,0)[bl]{{\Large $\mC$ with mixing}}}
\end{picture}
\mbox{\epsffile{fig24m.eps}}

\vspace*{5mm}

Figure~\thefigure
\end{center}
\end{figure}

\clearpage

\begin{figure}
\refstepcounter{figure}
\label{plsh}
\setlength{\unitlength}{1cm}
\begin{center}
\begin{picture}(16,0)
\put(7,-5.3){{\Large\bf $\mh$}}
\put(10,-3){{{\Large (a) \hspace*{3mm} no mixing}}}
\put(7,-15.3){{\Large\bf $\mh$}}
\put(10,-13){{{\Large (b) \hspace*{3mm} mixing}}}
\end{picture}
\mbox{\epsffile{fig7h.eps}}

\vspace{5mm}

Figure~\thefigure
\end{center}
\end{figure}

\clearpage

\begin{figure}
\refstepcounter{figure}
\label{plfsh}
\setlength{\unitlength}{1cm}
\begin{center}
\begin{picture}(16,0)
\put(8,-4){{\Large\bf $\mh$}}
\put(10,-3){{{\Large (a) \hspace*{3mm} no mixing}}}
\put(8,-14){{\Large\bf $\mh$}}
\put(10,-13){{{\Large (b) \hspace*{3mm} mixing}}}
\end{picture}
\mbox{\epsffile{fig24h.eps}}

\vspace*{5mm}

Figure~\thefigure
\end{center}
\end{figure}

\clearpage

\begin{figure}
\refstepcounter{figure}
\label{pllhe}
\mbox{\epsffile{figlh.eps}}
\setlength{\unitlength}{1cm}
\begin{center}
\begin{picture}(16,0)
\put(1.5,20){{{\Large (a) \hspace*{3mm} no mixing}}}
\put(1.5,10){{{\Large (b) \hspace*{3mm} mixing}}}
\end{picture}
\raisebox{10mm}{Figure~\thefigure}
\end{center}
\end{figure}

\clearpage

\begin{figure}
\refstepcounter{figure}
\label{figov}
\begin{center}
\mbox{\epsffile{figc1.eps}}

\vspace{5mm}

Figure~\thefigure
\end{center}
\end{figure}

\clearpage

\begin{figure}
\refstepcounter{figure}
\label{plli}
\begin{center}
\mbox{\epsffile{fig10.eps}}

\vspace*{5mm}

Figure~\thefigure
\setlength{\unitlength}{1cm}
\begin{picture}(16,0)
\put(12,16){{\Large (a)}}
\put(12,5.5){{\Large (b)}}
\end{picture}
\end{center}
\end{figure}

\clearpage

\begin{figure}
\refstepcounter{figure}
\label{pla}
\begin{center}
\mbox{\epsffile{fig7f.eps}}

\vspace{5mm}

Figure~\thefigure
\end{center}
\end{figure}

\clearpage

\begin{figure}
\refstepcounter{figure}
\label{plfa}
\begin{center}
\mbox{\epsffile{fig24f.eps}}
Figure~\thefigure
\setlength{\unitlength}{1cm}
\begin{picture}(16,0)
\put(12.5,17.5){{\Large (a)}}
\put(12.5,7){{\Large (b)}}
\end{picture}
\end{center}
\end{figure}

\clearpage

\begin{figure}
\refstepcounter{figure}
\label{fighv}
\begin{center}
\mbox{\epsffile{fighv.eps}}

\vspace{5mm}

Figure~\thefigure
\end{center}
\end{figure}

\end{document}